\documentclass[transaction]{IEEEtran}
\usepackage{graphicx}
\usepackage{epstopdf}
\usepackage[latin1]{inputenc}
\usepackage{amsmath,amssymb,amscd,latexsym,dsfont}
\usepackage[english]{babel}
\usepackage{tabularx,cite}
\usepackage{graphicx}
\usepackage{algpseudocode}
\usepackage{algorithm}
\usepackage{algorithmicx}
\usepackage{float}
\usepackage{multicol}
\usepackage{psfrag}
\usepackage{comment,psfrag,subfigure,enumerate}
\usepackage{color}
\usepackage{amsthm}
\usepackage{graphicx}
\usepackage{graphicx}
\usepackage{epstopdf}
\usepackage[latin1]{inputenc}
\usepackage{amsmath,amssymb,amscd,latexsym,dsfont}
\usepackage[english]{babel}
\usepackage{tabularx,cite}
\usepackage{graphicx}
\usepackage{algpseudocode}
\usepackage{algorithm}
\usepackage{algorithmicx}
\usepackage{float}
\usepackage{multicol}
\usepackage{mathtools}
\usepackage{psfrag}
\usepackage{comment,psfrag,subfigure,enumerate}
\usepackage{color}
\usepackage{amsthm}
\usepackage{inputenc}

\ifCLASSINFOpdf
\else
\fi
\begin{document}
%
\title{A Survey of NOMA: Current Status and Open Research Challenges}

\author{\IEEEauthorblockN{Behrooz Makki, \emph{Senior Member, IEEE}, Krishna Chitti, Ali Behravan and Mohamed-Slim Alouini, \emph{Fellow, IEEE}}
\thanks{Behrooz Makki, Krishna Chitti, and Ali Behravan are with Ericsson Research, Sweden, Email: \{behrooz.makki, krishna.chitti, ali.behravan\}@ericsson.com}
\thanks{ Mohamed-Slim Alouini is with the King Abdullah University of Science and Technology (KAUST), Thuwal, Makkah Province, Saudi Arabia, Email: slim.alouini@kaust.edu.sa.}
}
\maketitle

\begin{abstract}
Non-orthogonal multiple access (NOMA) has been considered as a study-item in 3GPP for 5G new radio (NR). However, it was decided not to continue with it as a work-item, and to leave it for possible use in beyond 5G. In this paper, we first review the discussions that ended in such decision. Particularly, we present simulation comparisons between the NOMA and multi-user multiple-input-multiple-output (MU-MIMO), where the possible gain of NOMA, compared to MU-MIMO, is negligible. Then, we propose a number of methods to reduce the implementation complexity and delay of both uplink (UL) and downlink (DL) NOMA-based transmission, as different ways to improve its efficiency. Here, particular attention is paid to reducing the receiver complexity, the cost of hybrid automatic repeat request as well as the user pairing complexity.
As demonstrated, different smart techniques can be applied to improve the energy efficiency and the end-to-end transmission delay of NOMA-based systems.
\end{abstract}


%
\IEEEpeerreviewmaketitle
\vspace{-2mm}
\section{Introduction}
The design of multiple access schemes is of interest in the cellular systems design. Here, the goal is to provide multiple user equipments (UEs) with radio resources in a spectrum-, cost- and complexity-efficient manner. In 1G-3G, frequency division multiple access (FDMA), TDMA (T: time) and CDMA (C: code) schemes have been introduced, respectively. Then, Long-Term Evolution (LTE) and LTE-Advanced developed orthogonal frequency division multiple access (OFDMA) and single-carrier (SC)-FDMA as orthogonal multiple access (OMA) schemes. Also, 5G new radio (NR) utilizes OFDMA waveform in both uplink (UL) and downlink (DL) transmission. Such orthogonal designs have the benefit that there is no mutual interference among UEs, leading to high system performance with simple receivers.

In the last few years, non-orthogonal multiple access (NOMA) has received considerable attention as a candidate multiple access technique for LTE, 5G and beyond 5G systems. With NOMA, multiple UEs are co-scheduled and share the same radio resources in time, frequency and/or code. Particularly, 3GPP has considered NOMA in different applications. For instance, NOMA has been introduced as an extension of the network-assisted interference cancellation and suppression (NAICS) for inter-cell interference (ICI) mitigation in LTE Release 12 \cite{refnoma1} as well as a study-item of LTE Release 13, under the name of DL multi-user superposition transmission (DMST) \cite{refnoma2}.

Different schemes have been proposed for NOMA including, power domain NOMA \cite{poornoma}, SCMA (SC: sparse code) \cite{7752784,8625325}, PDMA (PD: pattern division) \cite{7526461},  RSMA (RS: resource spread) \cite{RSMA}, multi-user shared access (MUSA) \cite{7504361},  IGMA (IG: interleave-grid) \cite{IGMA}, Welch-bound equality spread multiple access (WSMA) \cite{WSAM_ericsson,4907428},  and IDMA (ID: interleave-division) \cite{7974726}.  These techniques follow the
superposition principle and, along with differences in bit- and symbol-level NOMA implementation, the main difference among them is the UEs' signature design which is based on spreading, coding, scrambling, or interleaving distinctness.


Various fundamental results have been presented to determine the ultimate performance of NOMA in both DL \cite{8558537,7812683,7433470,7542118,poornoma} and UL \cite{7433470,7542118,7390209,7913656}, to incorporate the typical data transmission methods such as hybrid automatic repeat request (HARQ) to the cases using NOMA \cite{7501524,8408492,8335325}, to develop low-complexity UE pairing schemes \cite{7557079,8016604,7511620}, and to reduce the receiver complexity \cite{8423459,8630078,7752784}. As shown in these works, with proper parameter settings, NOMA has the potential to outperform the existing OMA techniques at the cost of receiver, UE pairing and coordination complexity. For these reasons, NOMA has been suggested as a possibility for data transmission in dense networks with a large number of UEs requesting for access such that there are not enough orthogonal resources to serve them in an OMA-based fashion. Particularly, in 2018, 3GPP considered a study-item to evaluate the benefits of NOMA and provide guidelines on whether NR should support (at least) UL NOMA, in addition to the OMA \cite{3gppfinalnoma,refnoma3}. However, due to the reasons that we explain in the following, it was decided not to continue with NOMA as a work-item, and to leave it for possible use in beyond 5G.

In this paper, we study the performance of NOMA in UL systems (in the meantime, most the discussions are applicable/easy-to-extend for DL transmission). The contributions of the paper are threefold:
\begin{itemize}
  \item We summarize the final conclusions presented in 3GPP Release 15 study-item on NOMA. Particularly, we present the discussions leading to the conclusion of not continuing with NOMA as a work-item.  Such conclusions provide guidelines for the researchers on how to improve the practicality of NOMA.
  \item We present link-level evaluation results to compare the performance of WSMA-based NOMA and multi-user multiple-input-multiple-output (MU-MIMO) in different conditions. Here, the results are presented for the cases with both ideal and non-ideal channel estimation. As we show, the relative performance gain of NOMA compared to MU-MIMO, in terms of block error rate (BLER), is not that large to motivate its implementation complexity.
  \item We demonstrate different techniques to reduce the implementation complexity of NOMA-based systems. Here, we concentrate on developing low-complexity schemes for UE pairing, receiver design and NOMA-HARQ, where simple methods can be applied to reduce the implementation complexity of NOMA remarkably. These results are interesting for academia because each of the proposed schemes can be extended and studied analytically in a separate technical paper.
\end{itemize}
As we demonstrate, different techniques can be applied to reduce the implementation complexity of NOMA. Moreover, there is a need to improve the spectral efficiency and the practicality of implementation, in order to have NOMA adopted by the industry.
\section{Performance Analysis}
In this section, we first present the principles of WSMA as an attractive spreading-based NOMA technique. Then, we compare the performance of WSMA NOMA with MU-MIMO and summarize the final conclusions presented in 3GPP Release 15 study-item on NOMA.

\subsection{WSMA-based NOMA}
WSMA is a spreading-based NOMA scheme \cite{R11802767}. Here, the key feature is to use non-orthogonal short spreading sequences with relatively low cross-correlation for distinguishing multiple users, and the spreading sequences are non-sparse. The WSMA spreading sequences are based on the Welch bound \cite{WSAM_ericsson,4907428}, the details of which are explained in the following.

Let us consider $K$ UEs and signals of dimension $L$. The focus here is limited to symbol-level NOMA where each UE is assigned a UE specific vector from a set of pre-designed vectors. These vectors jointly have certain correlation properties. Consider $K$ vectors,  $\{\mathbf{s}_k, k=1,\ldots,K\}$ called signature sequences (SS), such that each $\mathbf{s}_k$ is of the dimension $(L \times 1)$ and $||\mathbf{s}_k||_2=1,\forall k,$ where $||\mathbf{s}_k||_2\doteq \sum_{l=1}^{L}{s_{k,l}}^2$. Let $\mathbf{S}=[\mathbf{s}_1, \mathbf{s}_2, \cdots, \mathbf{s}_K]$, be the overall $(L \times K)$ signature matrix. {The factor $\frac{K}{L}$ is referred to as the overloading factor in the WSMA context. Since one of the objectives of NOMA is to support a higher user density, it is required to have $\frac{K}{L}>1$. However, beyond a certain  value of $\frac{K}{L}$, the system will be interference-limited.} Depending on the required correlation properties of
$\mathbf{S}$, a certain performance indicator (PI) is chosen and optimized for the generation of $\mathbf{S}$. One such PI is the total squared correlation ($\text{TSC}$) and is given as
\begin{equation}
\text{TSC} = \sum\limits^K_{i=1} \sum\limits^K_{j=1} |\mathbf{s}_i^\text{H} \mathbf{s}_j|^2,
\end{equation}
{where $(\cdot)^\text{H}$ denotes the Hermitian operator}. This scalar PI is lower-bounded by a value called the Welch bound (WB) and is given as \cite{WSAM_ericsson,4907428}
\begin{align}
\text{TSC}\ge \frac{K^2}{L}.
\end{align}
 On obtaining the optimal value of the chosen PI, $\text{TSC}$ in this case, the WB is satisfied with equality and the set $\mathbf{S}$ is called a Welch bound equality (WBE) set. The constituent SSs satisfy WB as an ensemble and not individually, so there exists several sets $\mathbf{S}$ with similar correlation properties satisfying the WB for the same optimal PI. Also, it is required to have a low correlation value, given as
$\rho_{ij}=|\mathbf{s}_i^H \mathbf{s}_j|$, between the constituent vectors of $\mathbf{S}$. The motivation to have $ \text{TSC}=\frac{K^2}{L}$ is that, at the equality, several performance metrics in the system, such as sum-capacity and sum- mean square error (MSE), are optimized simultaneously \cite{WSAM_ericsson,4907428}. This makes it an attractive option for multiple access implementation. Such optimization and SS generation are well understood in the context of interference avoidance techniques \cite[Chapter 2]{popescu2006interference}.

Other PIs that may be considered for the SS generation include the worst-case matrix coherence given as $\mu = \max_{i \neq j} \rho_{ij}$  and the minimum chordal distance $d_{\textrm{cord}}$ (for detailed mathematical definition see \cite{calderbank2015block}). Optimizing each PI separately will result in a set of SSs each with a different set of correlation properties. Each of these sets is a subset of the WBE set. The number of vectors in each set must be decided before the optimization of the respective PIs. As an example, optimizing $\text{TSC}$ will result in a WBE set whose constituent vectors may have unequal correlation among them. Similarly, optimizing the worst-case matrix coherence $\mu$ will also produce a WBE set but with an additional property that each constituent SS is equally correlated with every other SS in the set. Such a set is known as a Grassmann set or an equiangular set, and the optimization problem is often referred to as line-space packing problem \cite{tropp2005complex}.

At times, it may be required to have zero correlation between few vectors of the constituent SSs in the set. In that case, optimizing $d_{\textrm{cord}}$  is an attractive option. The optimization problem is then referred to as sub-space packing problem \cite{calderbank2015block}. Equations (\ref{eq:eqx1})-(\ref{eq:eqx3}) show the correlation properties of $\mathbf{S}$, each generated by optimizing a different PI, w.r.t the element-wise absolute value of the $(K \times K)$ Grammian matrix ($\mathbf{S}^H\mathbf{S}$) when the number of active UEs $K=4$. This $\mathbf{S}^H\mathbf{S}$ matrix is independent of the dimension $L$ and is given for different PIs as follows\footnote{WSMA is mainly designed for overloaded systems where $K>L$. However, for simplicity, (\ref{eq:eqx1})-(\ref{eq:eqx3}) show the Grammian matices w.r.t the mentioned PIs at $100\%$ overloading where $K=L$.}

\begin{equation}\label{eq:eqx1}
|\mathbf{S}^H\mathbf{S}|_{\text{TSC}} =
\begin{bmatrix}
1 & \rho_{12} & \rho_{13} & \rho_{14} \\
\rho_{12} & 1 & \rho_{23} & \rho_{24} \\
\rho_{13} & \rho_{23} & 1 & \rho_{34} \\
\rho_{14} & \rho_{24} & \rho_{34} & 1
\end{bmatrix},
\end{equation}

\begin{equation}\label{eq:eqx2}
|\mathbf{S}^H\mathbf{S}|_{\mu} =
\begin{bmatrix}
1 & \rho & \rho & \rho \\
\rho & 1 & \rho & \rho \\
\rho & \rho & 1 & \rho \\
\rho & \rho & \rho & 1
\end{bmatrix},
\end{equation}

\begin{equation}\label{eq:eqx3}
|\mathbf{S}^H\mathbf{S}|_{\textrm{d}_{\textrm{cord}}} =
\begin{bmatrix}
1 & 0 & \rho_{13} & \rho_{14} \\
0 & 1 & \rho_{23} & \rho_{24} \\
\rho_{13} & \rho_{23} & 1 & 0 \\
\rho_{14} & \rho_{24} & 0 & 1
\end{bmatrix}.
\end{equation}

\subsection{NOMA vs MU-MIMO}
\begin{figure}
\centering
\includegraphics[width=0.96\columnwidth]{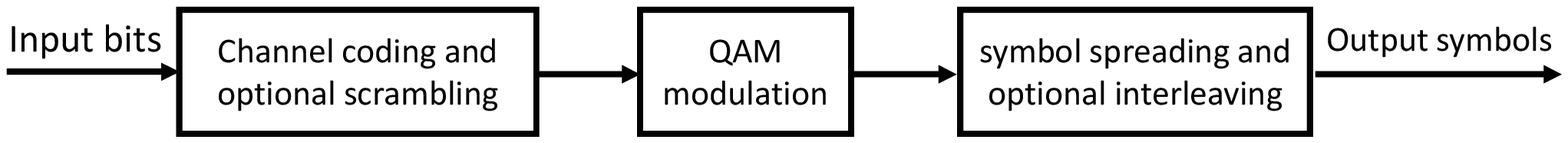}
\caption{Baseband transmitter implementation of WSMA-based NOMA at a user.}
\label{fig:fig_baseband_WSMA}
\end{figure}

A generalized block diagram of the baseband transmitter for NOMA implementation is shown in Fig. \ref{fig:fig_baseband_WSMA}. The information bits of a UE$_k$ are channel coded and then digitally modulated. For a bit-level NOMA implementation, the channel coded bits may be scrambled by a UE specific scrambling sequence and then digitally modulated. Symbol-level UE specific NOMA block appears after the quadrature amplitude modulation (QAM)-modulation block. Using WSMA, each incoming QAM-symbol $q_k$ is repeated $L$ times in a weighted manner by a UE assigned SS $\mathbf{s}_k$ to obtain an $(
L \times 1)$ output symbol vector $q_k \mathbf{s}_k$, i.e., a symbol spreading functionality.
The repeated symbols $q_k \mathbf{s}_k$ may optionally be interleaved to increase the randomness of the multiuser interference (MUI) to simplify the detector implementation. Usually, $\mathbf{S}$ is pre-generated in the system. To achieve collision-free multiple access, these SSs may be pre-assigned to the UEs, such that no two UEs have the same SS. Cooperation among the UEs may further improve the performance, but comes at an increased complexity and additional communication overhead among the UEs before the actual transmission to the base station (BS).

With $K$ NOMA transmitters, the received $(L \times 1)$ composite vector can be mathematically written as
\begin{equation}\label{eq:eqxx1}
\mathbf{y} = \sum\limits^K_{i=1} \mathbf{h}_k \odot \mathbf{s}_k \sqrt{p_k} q_k + \mathbf{z},
\end{equation}
where for a UE$_k$, $q_k$ is its QAM-symbol, $p_k$ is its transmit power, and $\mathbf{h}_k$ is the $(L \times 1)$ fading channel to the BS. Also, $\mathbf{z}$  is the $(L \times 1)$ zero mean AWGN vector, and $\odot$ is the element-wise multiplication. Symbol repetition at each UE may be performed in the frequency domain, but it can also be applied in the code domain.
With a scalar value $p_k$, the transmitter allocates the same power to all its incoming QAM symbols. This may be replaced by an $(L \times 1)$ per subcarrier power allocation power vector $\mathbf{p}_k$. Finally, note that (\ref{eq:eqxx1}) assumes that each UE is equipped with a single transmit antenna. However, it may be extended for higher number of transmit antennas, where each spatial layer may have its own SS.

With WSMA, each UE uses $L$ times more resources to transmit the same number of QAM-symbols. This may not be spectrally efficient. Hence it is important to overload the system, i.e., increase $K$ for a fixed $L$, to increase the sum-rate. This may lead to situations that require optimization of different metrics with conflicting interests.

The baseline system for comparison with NOMA could be an OMA setup when there are $K=L$ users. This is similar to $100\%$ overloading with SS matrix equal to identity matrix of size $K\times K$, and the UEs are scheduled over orthogonal resources.
The receiver, BS in this case, receives the composite signal and separates UEs in the frequency domain.

In another case, a baseline system for comparison could be based on MU-MIMO \cite{R11806248, R11808981}. In this case, both the NOMA and the MU-MIMO systems could be compared for the same number of users per RE. In addition to the frequency domain, the space domain provides additional degrees of freedom (DoF) to the BS. With multiple receive antennas at the BS, a joint space-frequency multiuser detector may be employed. The MU-MIMO system relies only on the spatial separation while NOMA has additional frequency domain, the assumed spreading domain, for UE separation. The same multiuser detector, with a little or no modification in implementation, may be used for both NOMA and MU-MIMO. For the MU-MIMO, an additional UE grouping and scheduling each group over orthogonal REs must also be considered for a fair comparison. Since increasing UE density is one of the NOMA objectives, a comparison for the maximum number of admissible UEs at a given target BLER may also be verified while comparing NOMA and MU-MIMO.

\begin{figure}
\centering
\includegraphics[width=0.96\columnwidth]{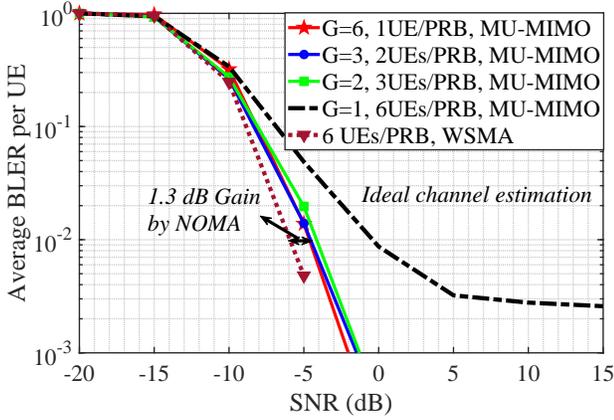}
\caption{NOMA vs MU-MIMO ($K=6,$ an ideal channel estimation).}
\label{fig:fig02}
\end{figure}

\begin{figure}
\centering
\includegraphics[width=0.96\columnwidth]{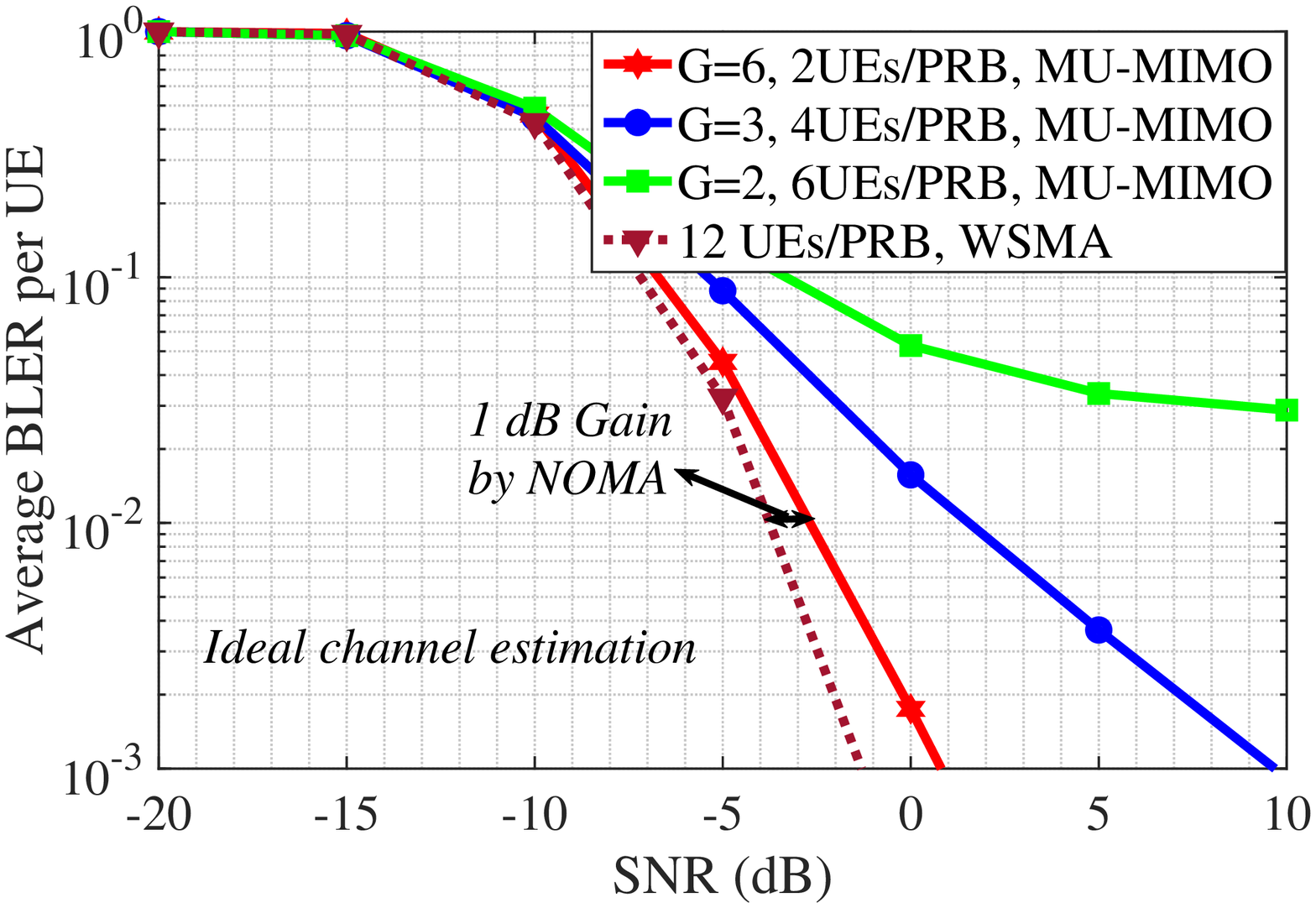}
\caption{NOMA vs MU-MIMO ($K=12$, an ideal channel estimation).}
\label{fig:fig03}
\end{figure}

Considering ideal channel estimation, Figs. \ref{fig:fig02} and \ref{fig:fig03}  show the link-level performance comparison of WSMA with MU-MIMO when the modulation is QPSK and the transport block size (TBS) is $20$ bytes. The carrier frequency is $700$ MHz and we assume that the channels follow the Tapped Delay Line (TDL-C) model \cite[Section 7.7.2]{TDL_model}. Note that TDL-C channel model may be used for simplified evaluation of non-line-of-sight (NLOS) communication.  The considered channel model for link-level simulations suffices the NOMA setup which usually targets a high user density coupled with low mobility and small delay spread values.

There are four receive antennas at the BS and each UE is equipped with a single antenna. It is assumed that each UE is transmitting with a unit power value over its allocated $6$ physical resource blocks (PRBs)  and $12$ data OFDM symbols. The detector at the BS is MMSE (M: minimum) based. With the spread length $4$, the codebook is based on the PI $\text{TSC}$. The channel encoding employed is the rate-matched LDPC code. There is no scrambling and interleaving at the transmitters. AWGN is assumed to have a unit variance. In Figs. \ref{fig:fig02} and \ref{fig:fig03}, the average BLER values per UE are shown for varying number of UEs $K=6$ and $K=12$, respectively. For a given $K$, to have the same number of UEs per PRB as in the case of WSMA, MU-MIMO divides the UEs into varying number of groups $G$ and varying number of UEs per group $N_\textrm{u}$ such that  $K=GN_\textrm{u}$.

From Figs. \ref{fig:fig02} and \ref{fig:fig03}, it can be observed that, for the assumed setup and various values of $G$, WSMA outperforms MU-MIMO, in terms of BLER, if ideal channel estimation is considered. There is also a saturation observed for MU-MIMO when it is heavily loaded. MU-MIMO systems are interference-limited, i.e., beyond a certain signal-to-noise ratio (SNR), an increase in the transmit power at each user may result in diminishing returns of the performance. A user's signal is drowned in the multiple access interference (MAI). With the overloading that NOMA targets, the available spatial DoF in MU-MIMO system are not sufficient to isolate the constituent signals from the received composite signal. This leads to the saturation in the BLER of MU-MIMO.

With NOMA, on the other hand, due to symbol repetition by the low correlation spreading, the energy per resource element (RE) on an average is reduced (note: the SS are unit norm). This ensures that users' signals perceive a lower MAI. This, however, comes at a possible reduced spectral efficiency, since each NOMA UE will consume $L$ times more REs than its MU-MIMO counterpart. This is very prominent at lower overloading factor, where the MU-MIMO outperforms NOMA. Hence a trade-off exists between the overloading and spectral efficiency. Nevertheless, with NOMA the error floors are lowered thereby providing a possibility to squeeze in more UEs per RE for the same target BLER as in MU-MIMO. However, as $G$ increases, MU-MIMO experiences saturation in lower BLERs and the difference between NOMA and MU-MIMO decreases. This is because by increasing $G$ in MU-MIMO, the  number of users per RE is reduced, leading to less multiuser interference for each user. More importantly, even with an ideal channel estimation, the relative performance gain of NOMA, compared to MU-MIMO, is negligible, and the relative performance gain decreases with the number of UEs. For instance, considering the parameter setting of Figs. \ref{fig:fig02} and \ref{fig:fig03} and BLER $10^{-2}$, NOMA-based data transmission reduces the required SNR, compared to MU-MIMO, only by $1.3$ and $1$ dB in the cases with $K=6$ and $K=12$ UEs, respectively. A definite advantage of having a higher UE density with NOMA is visible from Fig. 3.



\begin{figure}
\centering
\includegraphics[width=0.96\columnwidth]{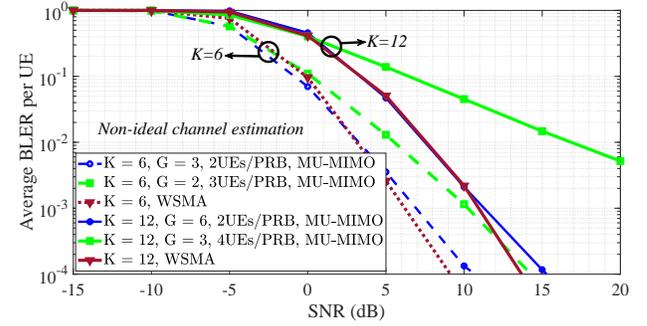}
\caption{NOMA vs MU-MIMO. $K=6$ and $K=12$, non-ideal channel estimation.}
\label{fig:fig04}
\end{figure}

Figure \ref{fig:fig04} compares WSMA and MU-MIMO for both $K=6$ and $K=12$, but with non-ideal channel estimation. A TDL-A channel model is assumed. At the UEs, the modulation is $16$-QAM and the TBS is $60$ bytes. As before, the BLER performance of MU-MIMO depends on its configuration, i.e., on the parameters $G$ and $N_{\text{u}}$. For $K=12$, the performance of MU-MIMO with $G=6$ and WSMA is very similar over a wide range of considered SNRs, with the latter outperforming the former only beyond a target BLER of $10^{-3}$.  When the setup is relatively less dense, i.e., $K=6$, a similar trend is observed between MU-MIMO with $G=3$ and WSMA. The target BLER beyond which WSMA performs better is now $10^{-2}$.  At lower SNR, less than $0$ dB, MU-MIMO has a better performance when compared to WSMA. This is also the case for MU-MIMO with $G=2$. These results do not indicate a possible advantage of MU-MIMO over WSMA and vice versa. However, considering different cases, the performance gain of NOMA, compared to MU-MIMO, is negligible.

\subsection{NOMA for beyond 5G}
During the NOMA Study in 3GPP for 5G NR, a large number of link-level simulations of transmission schemes and corresponding receivers were carried out \cite{3gppfinalnoma}. Particularly, 14 different companies, each with its own NOMA scheme, provided link-level results and studied the BLER for more than 35 cases. The link-level parameters were generally well aligned among companies, which enabled easy comparison between different methods. Moreover, all NOMA schemes, including those supported by Rel-15, performed similarly at link-level in key conditions. Then, 8 companies provided system-level simulation results, where in total 37 different sets of NOMA versus baseline results were provided. As opposed to the cases with link-level simulations, widely different parameter sets were used in the system-level simulations, with different baselines, making comparisons intractable. Here, the results have been presented for both synchronous and asynchronous operation models, while the main focus was on the synchronous operation.

According to the results presented during the 3GPP study-item, in ideal conditions, NOMA can be better or worse than MU-MIMO, depending on the number of UEs and simulation parameters. With realistic channel estimation in multipath, however, the relative perfromance gain of NOMA decreases, and MU-MIMO may outperform NOMA, depending on the parameter settings/channel model. No clear gain from NOMA over Rel-15 mechanisms were observed in all studied scenarios. Particularly, in a large number of conditions, link-level results from many companies show no gain over Rel-15 techniques and the system-level simulations do not show conclusive gain. In summary, in harmony with our results presented in Figs. \ref{fig:fig02}-\ref{fig:fig04}, it was hard to find worthwhile NOMA gains. This was the main reason that 3GPP decided not to continue with NOMA as a work-item, and leave it for beyond 5G where new use-cases with ultra-dense UEs may be motivating for NOMA.

\section{Reducing the Implementation Complexity}
One of the key challenges of NOMA is the implementation complexity, in different terms of UE pairing, signal decoding, CSI acquisition, etc.
This is specially because NOMA is useful in dense networks where the implementation complexity of the system increases rapidly with the number of UEs. This section proposes different techniques to reduce the implementation complexity of NOMA. These results are interesting because 1) they provide guidelines to use NOMA with relatively low complexity. Also, 2) each of the proposed schemes, which have been filed in patent applications, can be studied analytically by academia in a separate paper.
\begin{figure}
\vspace{-3mm}
\centering
  \includegraphics[width=.96\columnwidth]{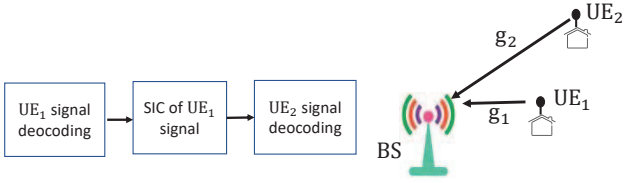}\\\vspace{-0mm}
\caption{UL NOMA. UEs with different channels qualities are paired and the BS performs SIC to decode the signals sequentially.}
\label{fig:fig_UL_NOMA}
\vspace{-0mm}
\end{figure}

For generality and in harmony with the discussions in 3GPP, we present the proposed schemes for UL NOMA. However, it is straightforward to extend our proposed approaches to the cases with DL transmission. We consider the cases with pairing a cell-center and a cell-edge UE, i.e., $\text{UE}_1$ and $\text{UE}_2$ in Fig. \ref{fig:fig_UL_NOMA}, respectively, with $g_1\ge g_2$ where $g_i$ denotes the channel gain in the $\text{UE}_i$-BS link. However, the discussions hold for arbitrary number of paired UEs. Moreover, for simplicity, we present the setups for the cases with power-domain NOMA and successive interference cancellation (SIC)-based receivers, while the same approaches are applicable for different types of NOMA-based data transmission/receivers. We concentrate on reducing the implementation complexity of the HARQ-based data transmission, UE pairing and receiver as follows.

\subsection{HARQ using NOMA}
Due to the CSI acquisition and UE pairing overhead, NOMA is of most interest in fairly static channels with no frequency hopping where channels remain constant for a number of packet transmissions. As a result, the network suffers from poor diversity. Also, NOMA is faced with error propagation problem where, if the receiver fails to decode a signal, its interference affects the decoding probability of all remaining signals which should be decoded sequentially. For these reasons, there may be a high probability for requiring multiple HARQ retransmissions leading to high end-to-end (E2E) packet transmission delay \cite{7501524,8408492,8335325}. The following schemes develop NOMA-HARQ protocols with low implementation complexity.
\subsubsection{Smart NOMA-HARQ \cite{patent13}}
\begin{figure}
\vspace{-3mm}
\centering
  \includegraphics[width=0.96\columnwidth]{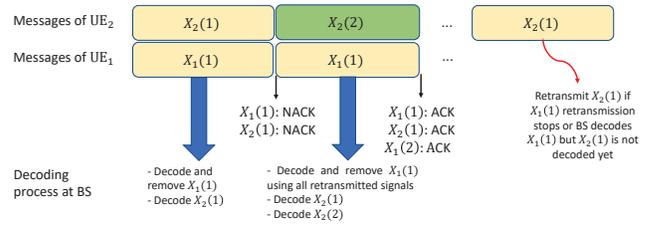}\\\vspace{-2mm}
\caption{Reducing the expected number of retransmissions in NOMA. If the BS fails to decode both signals, it asks for retransmission from only one of the UEs, while the other UE delays the retransmission. The retransmission gives the chance to decode the retransmitted signal. Then,  removing the interference, the BS can decode the other failed signal interference-free and with no need for retransmission.}
\label{fig:fig_IVD13}
\vspace{-0mm}
\end{figure}

Our proposed retransmission process is explained in Fig. \ref{fig:fig_IVD13}. Assume that in Slot 1 the BS can not decode correctly the signals of the UEs, i.e., $X_1(1)$ and $X_2(1)$. While buffering both failed signals, it asks $\text{UE}_2$ to delay the retransmission of the failed signal. Also, it asks $\text{UE}_1$ (resp. $\text{UE}_2$) to retransmit the failed signal $X_1(1)$ (resp. send a new signal $X_2(2)$) in Slot 2. At the end of Slot 2, the BS first combines the two interference-affected copies of $X_1(1)$ and decodes it using, e.g., maximum ratio combining (MRC). If the signal of $\text{UE}_1$ is correctly decoded, the BS has the chance to use SIC, remove $X_1(1)$ and decode both the failed and the new signals of $\text{UE}_2$, i.e., $X_2(1)$ and $X_2(2)$ received in Slots 1 and 2, respectively, interference-free and with no need for the retransmission from $\text{UE}_2$. Finally, $\text{UE}_2$ starts retransmitting $X_2(1)$ only if the retransmission of $\text{UE}_1$ stops (either because the maximum number of retransmissions is reached or the BS has correctly decoded the signal of $\text{UE}_1$) while the signal of $\text{UE}_2$ has not been decoded yet.

In this way, NOMA gives an opportunity to reduce the number of retransmissions, and improve the E2E throughput. Also, the fairness between the UEs increases because the required number of retransmissions of the cell-edge UE, i.e., $\text{UE}_2$ in Fig. \ref{fig:fig_IVD13}, decreases remarkably. The keys to enable such a setup are that 1) the BS should decode all buffered signals in each round and 2) it should inform the UEs about the appropriate retransmission times.

\subsubsection{Dynamic UE Pairing in NOMA-HARQ \cite{patent6}}
Here, the objective is to improve the performance gain of NOMA-HARQ by adding \emph{virtual} diversity into the network. In our proposed setup, depending on the message decoding status, different pairs of UEs may be considered for data transmission in different retransmission rounds. As an example, considering Fig. \ref{fig:fig_UL_NOMA}, assume that $\text{UE}_1$ and $\text{UE}_2$ with $g_1\ge g_2$ are paired and send their signals to the BS in a NOMA-based fashion. However, the BS fails to decode $X_1(1)$ (and with high probability $X_2(1)$). Then, in \cite{patent6}, we propose that in the retransmission(s) $\text{UE}_1$ can be paired with a new UE, namely, $\text{UE}_0$ with $g_0\ge g_1$. This is intuitively because a large portion of the SNR required for successful decoding of $X_1(1)$ has been provided in Round 1. Thus, although we have failed, we are very close to successful decoding and the signal can be correctly decoded  by a small boost in the retransmission round. Such a boost can be given by pairing $\text{UE}_1$ with $\text{UE}_0$ having a better channel to the BS. On the other hand, pairing $\text{UE}_0$ and $\text{UE}_1$ gives $\text{UE}_2$ the chance to use a separate resource block to retransmit its own signal interference-free. Also, although the channel coefficients are constant, pairing different UEs in successive rounds provides the BS with different SNR/SINR (I: interference) powers, i.e., diversity, which improves the performance of HARQ protocols considerably. In this way, the fairness between the UEs and the expected E2E packet transmission delay of the network are improved.

\subsubsection{Multiple Access Adaptation in Retransmissions \cite{patent9}}
\begin{figure}
\vspace{-3mm}
\centering
  \includegraphics[width=0.96\columnwidth]{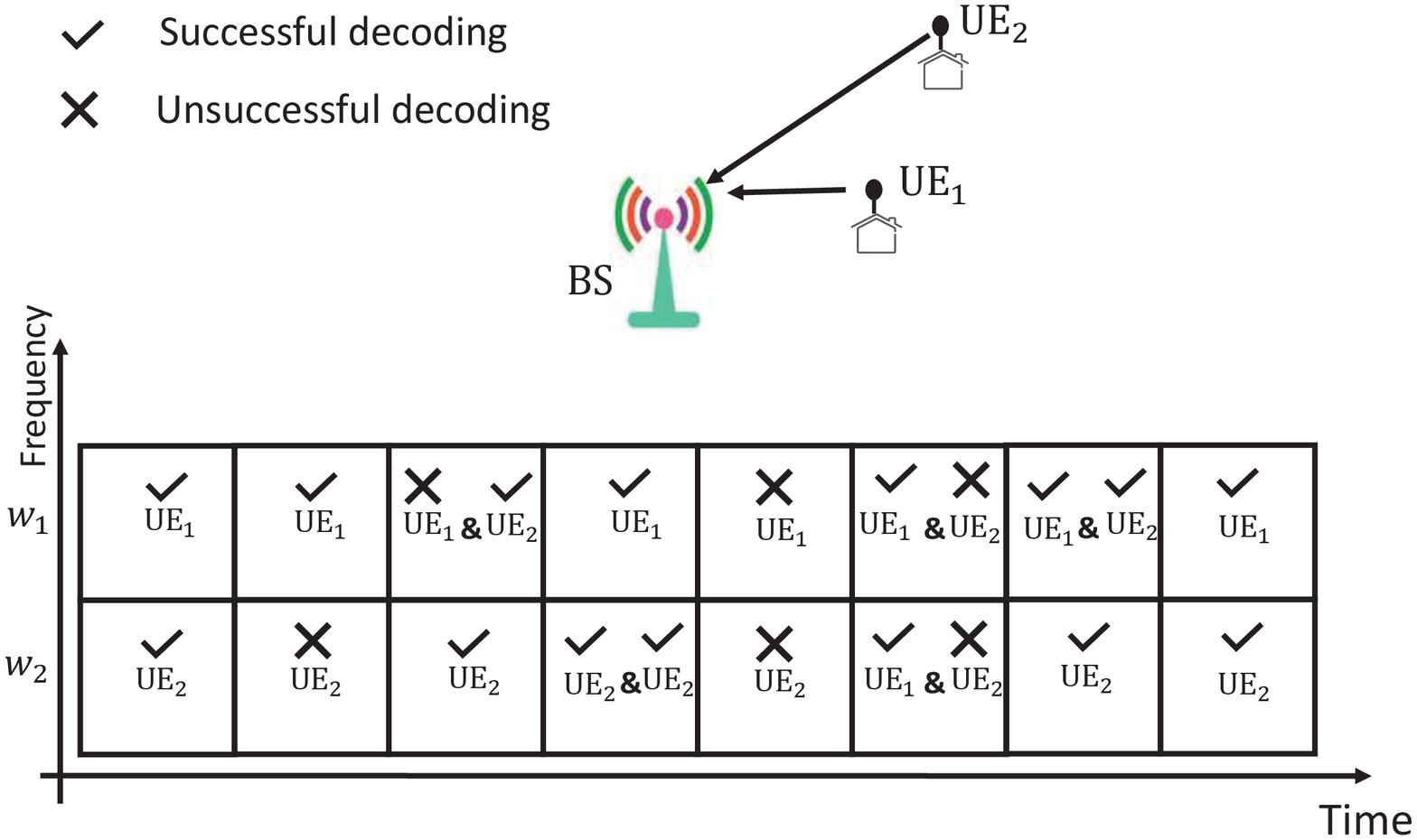}\\\vspace{-2mm}
\caption{Multiple access adaptation in different (re)transmission rounds. If the signal of an UE is not correctly decoded, it has the chance to reuse the spectrum resource of the other UE during retransmissions.}
\label{fig:fig_ivd9}
\vspace{-0mm}
\end{figure}

Our proposed scheme can be well explained in Fig. \ref{fig:fig_ivd9}. In our proposed setup, each UE starts data transmission in its own dedicated bandwidth in an OMA-based fashion. Then, if a UE's message is not correctly decoded in a time slot, in the following retransmission rounds it is allowed to reuse the bandwidth of the other UE as well. Let us denote the resource block at time $i$ and bandwidth $w_j$ by $B(i,w_j)$. As an example, consider time Slot 2 in Fig. \ref{fig:fig_ivd9} where the $\text{UE}_2$'s message is not correctly decoded while the message of $\text{UE}_1$ is successfully decoded by the BS. Then, in Slot 3, $\text{UE}_2$ uses both $w_1$ and $w_2$ to retransmit its message. On the other hand, $\text{UE}_1$ only uses $w_1$ to send a new message. Using, e.g., repetition time diversity (RTD) HARQ, in $B(3,w_1)$ and $B(3,w_2)$, $\text{UE}_2$ sends the same signal as in $B(2,w_1)$ and the BS decodes the message based on all three copies of the signal. An example method for message decoding is to first use SIC to decode the message of $\text{UE}_1$ in $B(3,w_1)$, then remove this message from the received signal in $B(3,w_1)$, and use MRC of the three copies of the $\text{UE}_2$'s signal to decode its message. Note that, even if the message of $\text{UE}_1$ is not correctly decoded in Slot 3, the BS can still perform, e.g., MRC of the three (two interference-free and one interference-affected) copies of the $\text{UE}_2$'s signal.

In this way, compared to the cases with conventional OMA techniques, using the adaptive multiple access scheme, along with HARQ, makes it possible to exploit the network/frequency diversity and increase the UEs' achievable rates. Moreover, our proposed scheme satisfies the tradeoff between the receiver complexity and the network reliability, and, compared to the state-of-the-art OMA-based systems, improves the service availability/the network reliability significantly. Finally, the proposed scheme improves the fairness between UEs and is useful in buffer-limited systems.

Finally, note that, while we presented the proposed NOMA-HARQ schemes for RTD (Type II) HARQ \cite{6804409}, the same approaches are applicable for other HARQ protocols as well.
\subsection{Simplifying the UE Pairing}

With NOMA, optimal UE pairing becomes challenging as the number of UEs increases, because it leads to huge CSI acquisition and feedback overhead as well as running complex optimization algorithms \cite{7557079,8016604,7511620}. For these reasons, we present low-complexity UE pairing schemes as follows.
\subsubsection{Rate-based UE Pairing \cite{patent4}}
Consider a dense network with $N$ UEs and $N_\text{c}$ time-frequency chunks where $N_\text{c}<N$, i.e., when the number of resources are not enough to serve all UEs in orthogonal resources. An optimal UE pairing algorithm needs to know all $N_\text{c}N$ channel coefficients and all $N$ rate demands of the UEs, making the whole system impractical as $N$ and/or $N_\text{c}$ increases. This is especially because a large portion of this information is used only for UE pairing and not for data transmission.

To limit the CSI requirement, in \cite{patent4}, we propose that the UE pairing is performed only based on the UEs rate demands and the probability of successful pairing. The proposed scheme is based on the following procedure:
\begin{itemize}
  \item Step 1: The BS asks all UEs to send their rate demands.
  \item Step 2: Receiving the UEs' rate demands and without knowing the instantaneous CSI, the BS finds the probability that two specific UEs can be successfully served through NOMA-based data transmission (see \cite{patent4} for the detail procedure of finding these probabilities).
  \item Step 3: If the probability of successful pairing for two specific UEs, i.e., the probability that the BS can correctly decode their signals, exceeds some predefined threshold, the BS assigns resources for UL transmission and asks those UEs to send pilots sequentially.
  \item Step 4: Using the received pilots from those paired UEs, the BS estimates the channel qualities in that specific resources, decides if the UEs can be paired and determines the appropriate power level of each UE such that their rate demands can be satisfied.
  \item Step 5: The BS informs the paired UEs about the power levels to use and sends synchronization signals such that their transmit timings are synchronized.
\end{itemize}

In this way, with our proposed scheme the CSI is acquired only if the BS estimates a high probability for successful UE pairing. This reduces the CSI overhead considerably, particularly in dense networks and/or in the cases with multiple antennas at the UEs. Finally, as we show in \cite{patent4}, to have the maximum number of successful paired UEs, the BS can initially consider the pairs with the highest and lowest rate demands. For instance, assume $r_1\ge\ldots\ge r_N$ where $r_i$ is the rate demand of $\text{UE}_i$. Then, an appropriate UE pairing approach would be $(r_1, r_N),(r_2, r_{N-1}),\dots.$

\subsubsection{UE Pairing in COMP-NOMA}
\begin{figure}
\vspace{-3mm}
\centering
  \includegraphics[width=0.96\columnwidth]{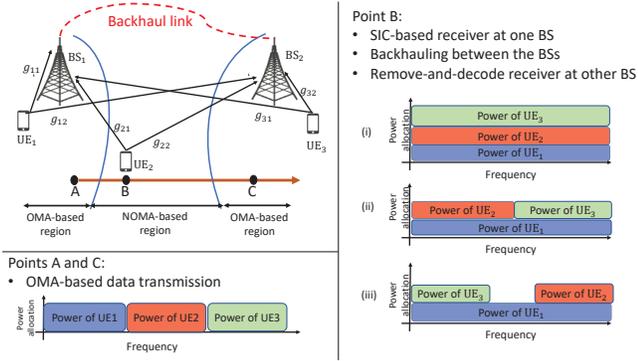}\\\vspace{-2mm}
\caption{UE pairing in COMP-NOMA. If a cell-edge UE can be successfully paired with a cell-center UE of one BS, it can be paired with each of the cell-center UEs of other BSs with SIC-based receiver only at one BS.}
\label{fig:fig_IVDx8}
\vspace{-0mm}
\end{figure}

The high-rate reliable backhaul links give the chance to simplify the UE pairing in coordinated multi-point (COMP) networks using NOMA. The idea can be well presented in Fig. \ref{fig:fig_IVDx8}. Here, depending on the UEs positions and rate demands, they may be served with different multiple access schemes. For instance, in Point A (resp. C) of Fig. \ref{fig:fig_IVDx8} where the channel $g_{21}$ (resp. $g_{22}$) experiences a good quality, UEs transmit in an OMA-based fashion and the BSs may use typical OMA-based receivers with no need for backhauling. In Point B, however, NOMA is used in a COMP-based fashion and the UEs may share spectrum. As an example, with $\text{UE}_2$ being in Point B, $\text{BS}_1$ may use SIC-based receiver to first decode-and-remove the message of  $\text{UE}_1$ and then decode the message of  $\text{UE}_2$. Then, using the backhaul link, $\text{BS}_2$ is informed about the message of $\text{UE}_2$ (and, possibly, $\text{UE}_1$) and, removing the interfering signal, it decodes the signal of $\text{UE}_3$ interference-free. Finally, as demonstrated in the figure, depending on the rate requirements and channel conditions, different NOMA-based transmissions with partial spectrum sharing can be considered in Point B, which affect the data transmission, backhauling and message decoding schemes correspondingly.

The advantages of the proposed scheme are: 1) SIC-based receiver is used only in $\text{BS}_1$. Also, 2) UE pairing algorithm can be run only in one of the BSs. That is, NOMA-based data transmission is used as long as at least one of the BSs can find a good pair for $\text{UE}_2$. Finally, 3) pairing $(\text{UE}_1$,$\text{UE}_2)$, $\text{BS}_2$ can consider each of its own cell-center UEs to be paired with them as long as the interference to $\text{BS}_1$ is not high. That is, $\text{BS}_2$ does not need to run advanced UE pairing algorithms.
\subsection{Receiver Adaptation}

Compared to OMA-based systems, the sequential decoding process of the BS may lead to large E2E transmission delay, as well as high receiver complexity/energy consumption \cite{8423459,8630078,7752784}. Therefore, it is beneficial to use the sequential decoding \emph{only if there is high probability for successful decoding.} This is the motivation for the scheme proposed in the following.

\begin{figure}
\vspace{-3mm}
\centering
  \includegraphics[width=0.96\columnwidth]{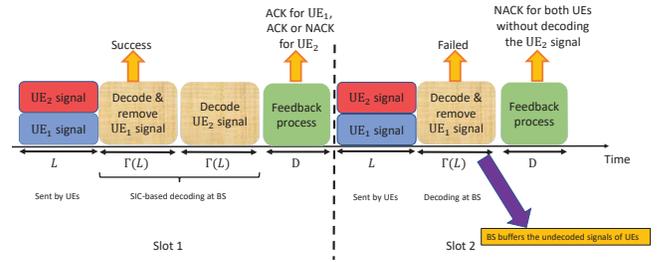}\\\vspace{-2mm}
\caption{Adapting the decoding scheme based on the estimated successful decoding probability.}
\label{fig:fig_IVD5}
\vspace{-0mm}
\end{figure}

Considering Fig. \ref{fig:fig_IVD5}, if the signal of $\text{UE}_1$ is correctly decoded, the BS continues in the typical SIC-based receiver scheme to first remove the signal of $\text{UE}_1$ and then decode the signal of $\text{UE}_2$ interference-free. On the other hand, if the BS fails to decode the signal of $\text{UE}_1$, it immediately sends NACKs for both UEs without decoding the $\text{UE}_2$'s signal. This is to save on the decoding delay and because there is low probability for successful decoding of the second signal without removing the first interfering signal. For instance, let us denote the decoding delay for decoding a codeword of length $L$ by $\Gamma(L).$ Then, as shown in Fig. \ref{fig:fig_IVD5}, the proposed scheme reduces the E2E delay by $\Gamma(L)$ in Slot 2. The BS buffers the undecoded signal of $\text{UE}_2$ for process in the next rounds of HARQ. Then, depending on the decoding approach of the BS and its corresponding decoding delay, in each time slot the UEs' data transmission is synchronized correspondingly.

In this way, the proposed setup reduces the implementation complexity considerably and improves the E2E throughput because the decoding scheme is adapted depending on the estimated probability of successful decoding. Particularly, an interested reader may follow the same method as in \cite{8525430} to study the E2E performance gain of the proposed scheme analytically. Finally, while we presented the setup for the cases with two UEs, it can be shown that the relative performance gain of the proposed scheme increases with the number of paired UEs.

\section{Conclusions}
In this paper, we studied the challenges and advantages of NOMA as a candidate technology in dense networks. As we showed through simulations and in harmony with the discussions in the 3GPP Release 15 study-item on NOMA, NOMA may or may not outperform the typical OMA-based schemes such as MU-MIMO, in terms of BLER. However, for the current use-case scenarios of interest, the relative performance gain of NOMA was not so much such that it could not convince the 3GPP to continue with it as a work-item. On the other hand, the unique properties of NOMA give the chance to develop different techniques reducing its implementation complexity, which may make it more suitable for practical implementation. Therefore, there is a need to improve the spectral efficiency and the practicality of implementation, in order to have NOMA adopted by the industry.

\vspace{-3mm}
\bibliographystyle{IEEEtran} 
\bibliography{masterMIMO}

\vfill

\end{document}